\documentclass[12pt,sec]{article}
\usepackage{epsfig}
\usepackage{axodraw}
\begin{document}

\begin{center}
{\LARGE D-wave Charmonium Production in $e^+e^-$ Annihilation at $\sqrt{s}=10.6$ GeV}\\[0.8cm]
{\large Li-Kun Hao$~^{(a)}$, Kui-Yong Liu$~^{(a)}$ and~Kuang-Ta Chao$~^{(b,a)}$}\\[0.5cm]
{\footnotesize (a)~Department of Physics, Peking University,
 Beijing 100871, People's Republic of China}

{\footnotesize (b)~China Center of Advanced Science and Technology
(World Laboratory), Beijing 100080, People's Republic of China}
\end{center}

\begin{abstract}

We calculate the D-wave charmonium production in $e^+e^-$
annihilation at BELLE and BABAR at $\sqrt{s}=10.6$ GeV using the
nonrelativistic QCD factorization formalism, including color
singlet and color octet contributions. We analyze the
contributions of various processes for $\delta_J(J=1,2)$
production. We find that both the color singlet and color octet
channels may give substantial contributions, and the production
rates are estimated to be $\sigma (\delta_1)\simeq 0.043-0.16pb$
and $\sigma (\delta_2)\simeq 0.094-0.29pb$, which , however, are
very sensitive to the choice of the color octet matrix elements of
the D-wave charmonium states. The measurement of D-wave charmonium
production at BABAR and BELLE in the future will be very helpful
to test the color octet mechanism and to determine the color octet
matrix elements.

PACS number(s): 12.40.Nn, 13.85.Ni, 14.40.Gx

\end{abstract}


Studies of heavy quarkonium production at high energies are hot
topics in recent years. In the conventional picture, the heavy
quarkonium production is described in the color-singlet
model\cite{cs}. In this model, it is assumed that the heavy quark
pair must be produced in a color-singlet state at short distance
with the same angular-momentum quantum number as the charmonium
which is eventually observed. However, with the recent Tevatron
data on high $p_T$ $J/\psi$ production, this color-singlet picture
for heavy quarkonium production has become questionable. The
observed cross sections are larger than the theoretical prediction
of the color-singlet model by a factor of about $30\sim
50$\cite{fa}. This is called the $J/\psi$ ($\psi'$) surplus
problem. On the theoretical side, the naive color-singlet model
may be supplanted by the nonrelativistic QCD (NRQCD) factorization
formalism \cite{bbl}, which allows the infrared safe calculation
of inclusive charmonium production and decay rates. In this
approach, the production process is factorized into short and long
distance parts, while the latter is associated with the
nonperturbative matrix elements of four-fermion operators. So, for
heavy quarkonium production, the quark-antiquark pair does not
need to be in the color-singlet state in the short distance
production stage, which is at the scale of $1/m_Q$ ($m_Q$ is the
heavy quark mass). At this stage, the color configuration other
than the singlet, i.e. the color-octet is allowed for the heavy
quark pair. The later situation for heavy quarkonium production is
called the color-octet mechanism. In this production mechanism,
heavy quark-antiquark pair is produced at short distances in a
color-octet state, and then hadronizes into a final state
quarkonium (physical state) nonperturbatively. With this
color-octet mechanism, one might explain the Tevatron data on the
surplus of $J/\psi $ and $\psi ^{\prime }$
production\cite{surplus,s1}.

Aside from the $J/\psi$ production at Tevatron, various heavy
quarkonium production processes have been studied to test the
color octet mechanism. Among them, the charmonium production in
$e^+e^-$ annihilation is particularly interesting, since in this
process, the parton structure is simpler, and there is no higher
twist effect, so the theoretical uncertainty is smaller and it can
even be used to extract the color-octet matrix elements. $J/\psi$
production in $e^+e^-$ annihilation process has been investigated
within the color-singlet model \cite{cm2,cm3,cm4,cm5,cm6,cm7} and
color-octet model\cite{om1,om2,ko}. Recently, BABAR\cite{babar}
and BELLE\cite{belle} have measured the direct $J/\psi$ production
in $e^+e^-$ annihilation at $\sqrt{s}=10.6 GeV$. The total cross
section and the angular distribution seem to favor the NRQCD
calculation over the color-singlet model\cite{babar}, but some
issues still remain. The P-wave charmonium $\chi_{cJ}$ production
in $e^+e^-$ annihilation has been discussed in \cite{xcprod}. The
total $\chi_{c1,2}$ cross sections are dominated by color-octet
process, because the C-parity suppresses the process
$e^+e^-\rightarrow c\bar{c}gg$, which dominates the $J/\psi$
production in the color singlet sector. In order to further test
the color octet mechanism, in this paper we study the D-wave
charmonium production in $e^+e^-$ annihilation. As pointed out by
the authors in \cite{qiao} and \cite{yuan}, due to the color octet
mechanism the D-wave charmonium production rates could be as large
as $J/\psi$ or $\psi'$ at the Tevatron, and could be enhanced by
two orders of magnitude at the fixed target experiments.
Therefore, it would also be interesting to study the D-wave
charmonium production in $e^+e^-$ annihilation.

We would like to mention that among the three spin-triplet $D$
wave charmonium states the ${}^3D_2$ is the most promising
candidate to discover. Its mass is predicted to be in the range of
$3.81-3.84GeV$ in the potential model calculation \cite{isgur},
which is above the $D\bar{D}$ threshold but below the $D\bar{D}^*$
threshold. The state ${}^3D_2(J^{PC}=2^{--})$ is forbidden to
decay into $D\bar{D}$ because of the parity conservation. So this
state is predicted to have a narrow decay width of $300-400 keV$,
and we may easily tag this state through the dominant decay
channels, e.g. the $E1$ transitions into $\chi_{cJ=1,2}$ states,
and the hadronic transition into $J/\psi+\pi\pi$(see \cite{qiao}
and \cite{yuan} for detailed discussions). For the $^3D_1$ state
$\psi''$(3770), it is just above the $ D\bar{D}$ threshold and
thus has a quite narrow width of about $23 MeV$ and may be tagged
via $\psi''\rightarrow D\bar{D}$. As for the
$^3D_3(J^{PC}=3^{--})$ state, its mass is far above the allowed
decay channel $ D\bar{D}$ threshold and it has more decay modes,
therefore it is expected to be a wide resonance and difficult to
observe. So, phenomenologically we are only interested in the
${}^3D_2$ as well as $^3D_1$ states, which will be discussed
below.

The power of the NRQCD formalism stems from the fact that
factorization formulae for observables are expansions in the small
parameter $v$ ,where $v$ is the average relative velocity of the
heavy quark and antiquark in quarkonium bound state. For
charmonium $v^2\sim 0.3$ , and for bottomonium $v^2\sim 0.1$ .
NRQCD velocity-scaling rules\cite{glcuk} allow us to estimate the
relative size of various NRQCD matrix elements. This information,
along with the dependence of the short-distance coefficients on
$\alpha_s$, permits us to decide which terms must be retained in
expressions for observables to reach a given level of accuracy. At
low orders, factorization formulas involve only a few matrix
elements, so several observables can be related by a small set of
parameters.

In NRQCD the Fock state expansion for the physical D-wave
charmonium (denoted by $\delta_J$) is
\begin{eqnarray}
\label{expandfock}
\nonumber
|\delta_J\rangle
&&=O(1)|c\bar{c}({}^{3}D_{J},\underline{1}) \rangle\\
\nonumber
&& +O(v)|c\bar{c}({}^{3}P_{J'},\underline{8})g\rangle\\
&& +O(v^2)|c\bar{c}({}^{3}S_1,\underline{8}~ or~
\underline{1})gg\rangle+\cdots.
\end{eqnarray}
The striking feature of this expansion is that although the
amplitudes of these Fock states in the expansion are different in
$v$ the contributions to the production rates of all these terms
are essentially of the same order of $v^7$. This is very different
from the case of $J/\psi$ and $\psi'$.

According to the NRQCD  factorization formalism\cite{bbl}, the production process
$e^+e^-\rightarrow\delta_J+X$ can be expressed as the following form,
\begin{equation}
\label{xs}
d\sigma(e^+e^-\rightarrow\delta_J+X)=\sum\limits_n F(e^+e^-\rightarrow n+X)\langle {\cal
O}_n^{\delta_J}\rangle .
\end{equation}
Here, $n$ denotes the $c\bar c$ pair configuration in the
expansion terms of Eq. (\ref{expandfock}) (including angular
momentum $^{2S+1}L_J$ and color index $\underline{1}$ or
$\underline{8}$). $F(e^+e^-\rightarrow n+X)$ is the short distance
coefficient for the subprocess $e^+e^-\rightarrow n+X$. $\langle
{\cal O}_n^{\delta_J}\rangle $ is the long distance
non-perturbative matrix element which represents the probability
of the $c \bar c$ pair in $n$ configuration evolving into the
physical state $\delta_J$. The short distance coefficient $F$ can
be calculated by using perturbative QCD in powers of $\alpha_s$.
The long distance matrix elements are still not available from the
first principles at present. However, the relative importance of
the contributions from different terms in Eq. (\ref{xs}) can be
estimated by using the NRQCD velocity scaling rules.

\begin{figure}
\centering
\begin{center}
\begin{picture}(200,110)(0,0)
\Text(90,10)[c]{(a)} \Text(80,70)[l]{$\gamma^*$}
\Text(20,30)[l]{$e^-$} \Text(20,95)[l]{$e^+$}
\Photon(50,60)(110,60){2.5}{6} \ArrowLine(0,10)(50,60)
\ArrowLine(50,60)(0,110) \ArrowLine(110,60)(125,70)
\ArrowLine(125,70)(170,70) \GOval(170,60)(10,4)(0){0.5}
\ArrowLine(170,50)(125,50) \ArrowLine(125,50)(110,60)
\Gluon(125,50)(175,10){2.5}{6} \Gluon(125,70)(175,110){2.5}{6}
\end{picture}\\
\vskip 8mm
\begin{picture}(200,110)(0,0)
\Text(90,10)[c]{(b)} \Text(80,70)[l]{$\gamma^*$}
\Text(20,30)[l]{$e^-$} \Text(20,95)[l]{$e^+$}
\Photon(50,60)(110,60){2.5}{6} \ArrowLine(0,10)(50,60)
\ArrowLine(50,60)(0,110) \ArrowLine(175,10)(110,60)
\Gluon(120,70)(155,70){2.5}{3} \ArrowLine(110,60)(170,110)
\ArrowLine(180,96)(152,70) \ArrowLine(152,70)(184,45)
\GOval(175,103)(9,4)(45){0.5}
\end{picture}\\
\vskip 8mm
\begin{picture}(200,110)(0,0)
\Text(90,10)[c]{(c)} \Text(80,70)[l]{$\gamma^*$}
\Text(20,30)[l]{$e^-$} \Text(20,95)[l]{$e^+$}
\Photon(50,60)(110,60){2.5}{6} \ArrowLine(0,10)(50,60)
\ArrowLine(50,60)(0,110) \ArrowLine(125,50)(110,60)
\ArrowLine(110,60)(170,110) \ArrowLine(180,96)(125,50)
\GOval(175,103)(9,4)(45){0.5} \Gluon(125,50)(175,10){2.5}{6}
\end{picture}\\
\end{center}
\caption{The main Feynman diagrams for the production of
$\delta_J$ in $e^+e^-$ annihilation}
\end{figure}
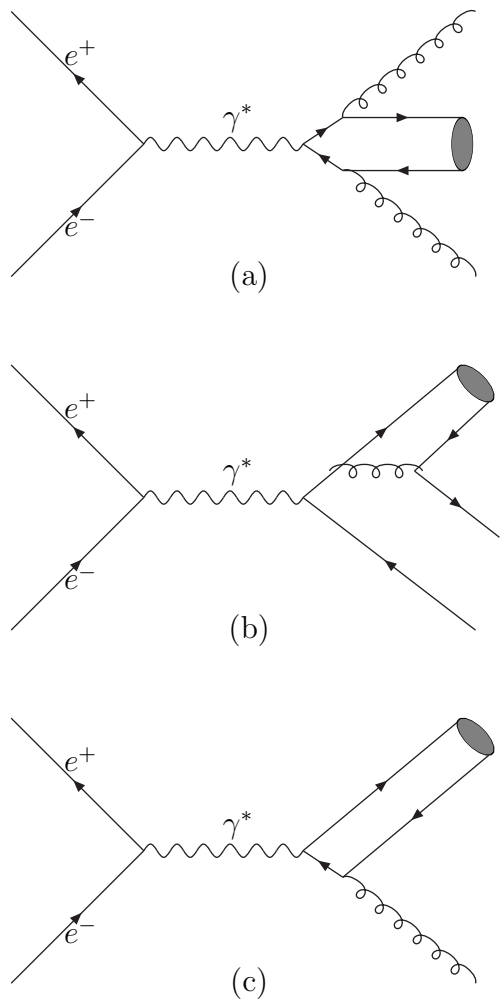

The main Feynman diagrams for the production of $\delta_J$ in
$e^+e^-$ annihilation are shown in Fig.1, where (a) and (b) are
the color singlet processes, and (c) is the color octet process.

For the color-singlet process in Fig.1(a),
\begin{eqnarray}
\label{ls} e^+e^-\rightarrow\gamma^*\rightarrow
c\bar{c}[{}^3D_J,\underline{1}]+g+g,
\end{eqnarray}

With the spin-triplet case where $J=1,~2,~3$, we use explicit
Clebsch-Gordan coefficients, and get the following relations for
the three cases.
\begin{eqnarray}
\nonumber \sum_{sm}\langle
1J_z|1s2m\rangle\epsilon_{\alpha\beta}^{(m)} \epsilon_\rho^{(s)}
&=&-[{3\over 20}]^{1/2}[(g_{\alpha\rho}-\frac{p_\alpha
p_\rho}{4M_c^2})\epsilon_\beta^{(J_z)}+(g_{\beta\rho}-\frac{p_\beta
p_\rho}{4M_c^2})\epsilon_\alpha^{(J_z)}\\
&~&-{2\over 3}(g_{\alpha\beta}-\frac{p_\alpha
p_\beta}{4M_c^2})\epsilon_\rho^{(J_z)}],\\
\sum_{sm}\langle 2J_z|1s2m\rangle\epsilon_{\alpha\beta}^{(m)}
\epsilon_\rho^{(s)}
&=&\frac{i}{2\sqrt{6}M_c}(\epsilon_{\alpha\sigma}^{(J_z)}
\epsilon_{\tau\beta\rho\sigma^{\prime}}p^\tau
g^{\sigma\sigma^\prime}+
\epsilon_{\beta\sigma}^{(J_z)}\epsilon_{\tau\alpha\rho\sigma^{\prime}}
p^\tau g^{\sigma\sigma^\prime}),\\
\sum_{sm}\langle 3J_z|1s2m\rangle\epsilon_{\alpha\beta}^{(m)}
\epsilon_\rho^{(s)} &=&\epsilon_{\alpha\beta\rho}^{(J_z)}.
\end{eqnarray}
Here, $\epsilon_\alpha$, $\epsilon_{\alpha\beta}$,
$\epsilon_{\alpha\beta\rho}$ are $J=1,2,3$ polarization tensors
which obey the projection relations
\begin{eqnarray}
\sum_m \epsilon_\alpha^{(m)}\epsilon_\beta^{(m)}
&=&(-g_{\alpha\beta}+\frac{p_\alpha p_\beta}{4M_c^2})
\equiv{\cal P}_{\alpha\beta},\\
\sum_m
\epsilon_{\alpha\beta}^{(m)}\epsilon_{\alpha^\prime\beta^\prime}^{(m)}
&=&{1\over 2}[{\cal P}_{\alpha\alpha^\prime}{\cal
P}_{\beta\beta^\prime}+ {\cal P}_{\alpha\beta^\prime}{\cal
P}_{\beta\alpha^\prime}]
-{1\over 3}{\cal P}_{\alpha\beta}{\cal P}_{\alpha^\prime\beta^\prime},\\
\nonumber \sum_m \epsilon_{\alpha\beta\rho}^{(m)}
\epsilon_{\alpha^\prime\beta^\prime\rho^\prime}^{(m)} &=&{1\over
6}({\cal P}_{\alpha\alpha^\prime} {\cal
P}_{\beta\beta^\prime}{\cal P}_{\rho\rho^\prime} +{\cal
P}_{\alpha\alpha^\prime}{\cal P}_{\beta\rho^\prime} {\cal
P}_{\beta\rho^\prime} +{\cal P}_{\alpha\beta^\prime}{\cal
P}_{\beta\alpha^\prime}
{\cal P}_{\rho\rho^\prime}\\
\nonumber &~&~~~~+{\cal P}_{\alpha\beta^\prime}{\cal
P}_{\beta\rho^\prime} {\cal P}_{\rho\alpha^\prime} +{\cal
P}_{\alpha\rho^\prime}{\cal P}_{\beta\beta^\prime} {\cal
P}_{\rho\alpha^\prime} +{\cal P}_{\alpha\rho^\prime}{\cal
P}_{\beta\alpha^\prime}
{\cal P}_{\rho\beta^\prime})\\
\nonumber &-&{1\over 15}({\cal P}_{\alpha\beta} {\cal
P}_{\rho\alpha^\prime}{\cal P}_{\beta^\prime\rho^\prime} +{\cal
P}_{\alpha\beta}{\cal P}_{\rho\beta^\prime} {\cal
P}_{\alpha^\prime\rho^\prime} +{\cal P}_{\alpha\beta}{\cal
P}_{\rho\rho^\prime}
{\cal P}_{\alpha^\prime\beta^\prime}\\
\nonumber &~&~~~~+{\cal P}_{\alpha\rho}{\cal
P}_{\beta\alpha^\prime} {\cal P}_{\beta^\prime\rho^\prime} +{\cal
P}_{\alpha\rho}{\cal P}_{\beta\beta^\prime} {\cal
P}_{\alpha^\prime\rho^\prime} +{\cal P}_{\alpha\rho}{\cal
P}_{\beta\rho^\prime}
{\cal P}_{\alpha^\prime\beta^\prime}\\
&~&~~~~+{\cal P}_{\beta\rho}{\cal P}_{\alpha\alpha^\prime} {\cal
P}_{\beta^\prime\rho^\prime} +{\cal P}_{\beta\rho}{\cal
P}_{\alpha\beta^\prime} {\cal P}_{\alpha^\prime\rho^\prime} +{\cal
P}_{\beta\rho}{\cal P}_{\alpha\rho^\prime} {\cal
P}_{\alpha^\prime\beta^\prime}).
\end{eqnarray}

We choose the parameters:
\begin{eqnarray}
\label{dtp} \nonumber
m_c=1.5GeV,  \alpha_s(2m_c)=0.26,\\
|R''_D(0)|^2=0.015GeV^7
\end{eqnarray}

For the color-singlet matrix element, we use
\begin{eqnarray}
\label{sme} \langle O_1^{\delta_J}({}^3D_J)\rangle
=\frac{15(2J+1)N_c}{4\pi}|R''_D(0)|^2.
\end{eqnarray}

Because these color-singlet processes have the infrared divergence
involved, we introduce an infrared cutoff, which can be set to
$m_c$ on the energy of the outgoing gluons in the quarkonium rest
frame. For these color-singlet contributions, we finally obtain
\begin{eqnarray}
\label{mp} \nonumber
\sigma_{singlet}(\delta_1)&=&0.027pb\\
\sigma_{singlet}(\delta_2)&=&0.067pb
\end{eqnarray}

For the color singlet process in Fig.1(b), we have estimated it in
the fragmentation limit, and found that its contribution to the
$\delta_1$ production cross section is smaller by a factor of
about 60 than that for the $J/\psi$ production cross section which
is about $0.1\sim 0.2pb$. This is consistent with the result given
in\cite{fragm}. So the contribution of Fig.1(b) to the $\delta_1$
production cross section is only $0.002\sim 0.004pb$, to
$\delta_2$ is $0.003\sim 0.006pb$, therefore is negligible.

We now discuss the color octet processes in Fig.1(c)
\begin{eqnarray}
\label{lo} e^+e^-\rightarrow\gamma^*\rightarrow
g+c\bar{c}[{}^{2S+1}L_J,\underline{8}],
\end{eqnarray}
we readily have
\begin{eqnarray}
\label{cslo} \sigma(e^+e^-\rightarrow\delta_J+g)=C_S\langle
O_8^{\delta_J}({}^1S_0)\rangle+C_P\langle
O_8^{\delta_J}({}^3P_0)\rangle,
\end{eqnarray}
with
\begin{eqnarray}
\label{loc}
C_s&=&\frac{64\pi^2e_c^2\alpha^2\alpha_s}{3}\frac{1-r}{s^2m},\\
C_p&=&\frac{256\pi^2e_c^2\alpha^2\alpha_s}{9s^2m^3}[\frac{(1-3r)^2}{1-r}+\frac{6(1+3)}{1-r}+\frac{2(1+3r+6r^2)}{1-r}],
\end{eqnarray}
where $r=m^2/s$, $m$ is the mass of $\delta_J$, and $s$ is the
$e^+e^-$ $c.m.$ energy squared. Here we have used the approximate
heavy quark spin symmetry relations
\begin{eqnarray}
\label{ssy} \langle O_8^{\delta_J}({}^3P_J)\rangle\approx
(2J+1)\langle O_8^{\delta_J}({}^3P_0)\rangle
\end{eqnarray}
Then we calculated the cross sections of $\delta_J$ production.
The cross section is
\begin{eqnarray}
\label{sdt} \sigma(e^+e^-\rightarrow\delta_J+X)= C_0\langle
O_{1,8}^{\delta_J}({}^3S_1)\rangle +C_1\langle
O_8^{\delta_J}({}^3P_0)\rangle +C_2\langle
O_1^{\delta_J}({}^3D_J)\rangle
\end{eqnarray}
Using the NRQCD velocity scaling rules, we roughly estimate the
$\delta_J$ color octet matrix elements by relating them to the
$\psi'$ color octet matrix elements:
\begin{eqnarray}
\label{ome}
\langle
O_8^{\delta_J}({}^3S_1)\rangle\approx\frac{(2J+1)}{5}\langle
O_8^{\psi'}({}^3S_1)\rangle
\simeq\frac{(2J+1)}{5}\times 4.6\times 10^{-3} GeV^3\\
\langle
O_8^{\delta_J}({}^3P_0)\rangle\approx\frac{(2J+1)}{5}\langle
O_8^{\psi'}({}^3P_0)\rangle \simeq\frac{(2J+1)}{5}\times 1.3\times
10^{-3} GeV^5
\end{eqnarray}
Using those color-octet matrix elements, we get
\begin{eqnarray}
\sigma_{octet}(\delta_1)&\simeq&0.016pb\\
\sigma_{octet}(\delta_2)&\simeq&0.027pb
\end{eqnarray}
Then we get the total cross section:
\begin{eqnarray}
\label{cs0}
\nonumber
\sigma(e^+e^-\rightarrow\gamma^*\rightarrow\delta_1+X)\simeq0.043pb\\
\sigma(e^+e^-\rightarrow\gamma^*\rightarrow\delta_2+X)\simeq0.094pb
\end{eqnarray}

However, if we use a more radical choice of the $\delta_J$ color
octet matrix elements by relating them to the $J/\psi$ color octet
matrix elements, we would get much larger values for the
$\delta_J$ production cross sections,
$\sigma(\delta_1)\simeq0.16pb$ and $\sigma(\delta_2)\simeq0.29pb$,
because the $J/\psi$ color octet matrix elements are much larger
than the $\psi'$ color octet matrix elements($\langle
O_8^{J/\psi}({}^3P_0)\rangle\simeq1.1\times 10^{-2} GeV^5$,
elements($\langle O_8^{\psi'}({}^3P_0)\rangle\simeq1.3\times
10^{-3} GeV^5$). Therefore the measurement of $\delta_J$
production rates in the future will be very helpful to determine
the values of the $\delta_J$ color octet matrix elements.

In summary, we have estimated the D-wave charmonium
$\delta_J(J=1,2)$ production rates in $e^+e^-$ annihilation at
$\sqrt{s}=10.6$ GeV. We find that both color singlet and color
octet may give substantial contributions to the production cross
sections, of which the values are sensitive to the choice of the
$\delta_J$ color octet matrix elements. We hope that these results
can be tested with higher statistic $e^+e^-$ annihilation data in
the future at BABAR and BELLE. For $\delta_2$, its branching
fraction of decay mode $J/\psi\pi^+\pi^-$ is estimated to be
$B(\delta_2\rightarrow J/\psi\pi^+\pi^-)\simeq 0.12$\cite{brr},
which is only smaller than that of $B(\psi'\rightarrow
J/\psi\pi^+\pi^-)=0.324\pm0.026$ by a factor of 3. We compare the
predicted production rate of $2^{--}$ D-wave charmonium with that
of $\psi'$. With the integrated luminosities of about $30 fb^{-1}$
at $\sqrt{s}=10.6$ GeV, BELLE gives $\sigma(e^+e^-\rightarrow
\psi'+X)\simeq 0.67\pm 0.09 pb$ with $143\pm 19$ $\psi'$ events
decaying to $J/\psi\pi^+\pi^-$ \cite{belle}. As a rough estimate,
if we choose $\sigma(\delta_2)=0.094pb$, about 7 events of
$2^{--}$ state decaying to $J/\psi\pi^+\pi^-$ will be detected,
and for $\sigma(\delta_2)=0.29pb$ (with larger values for the
color-octet matrix elements), it is about 23 events. With more
data available at the $B$ Factories in the future, it will be
possible to detect this $2^{--}$ D-wave charmonium state
(especially in the latter case). The $^3D_1$ $c\bar{c}$ state
$\psi''(3770)$ could also be detected via $\psi''\rightarrow
D\bar{D}$ decay.
\\

\noindent \textbf{Acknowledgments}

This work was supported in part by the National Natural Science
Foundation of China, and the Education Ministry of China.

\end{document}